\documentstyle[11pt,newpasp,twoside,epsf]{article}
\def\edcomment#1{\iffalse\marginpar{\raggedright\sl#1\/}\else\relax\fi}
\reversemarginpar

\begin{document}
\title{Polarized Microwave Radiation from Dust}
 \author{A. Lazarian}
\affil{Department of Astronomy, University of Wisconsin, Madison, WI 53706}

\begin{abstract}
Observations of cosmic microwave background in the range
10-90~GHz have revealed an anomalous foreground component well
correlated with 12~$\mu$m, 60~$\mu$m and 100~$\mu$m emission
from interstellar dust. As the recent cross-correlation analysis of 
WHAM H$\alpha$ maps with the Tenerife 10 and 15~GHz maps supports
an earlier conclusion that the emission does not arise from free-free 
radiation, the interstellar dust origin of it is left as the only
suspect. Two competing models of this emission exist. The more
favored at the moment is the spinning dust model, the other
is the model that uses grains with strong magnetic response. 
In the spinning dust model the emission
arises from rapid rotation of ultrasmall grains that have dipole moments, 
while in the
other model magnetic grains emit due to thermal vibrations
of magnetic dipoles. Both models predict the emission to be partially
polarized and this emission can seriously interfere with the CMB
polarization measurements. 
We discuss observational signatures that can be used to distinguish
and eventually filter out the polarized component of the microwave
dust radiation.  
\end{abstract}

\section{Introduction}

Diffuse Galactic microwave emission carries important information on 
the fundamental
properties of interstellar medium, but it also interferes with the
Cosmic Microwave Background (CMB) experiments 
(see Tegmark et al. 1999 and references
therein). Polarization of the CMB provides information about the Universe
that is not contained in the temperature data (see Prunet \& Lazarian
1999, Davis \& Wilkinson 1999) and a number of groups
around the world (see Table~1 in Staggs et al. 1999) work hard to determine
the CMB polarization. In view of this work, the issue of determining the
degree of Galactic foreground polarization becomes vital.

Microwave emission from Galactic dust has been recently identified as
an important component of Galactic microwave foreground (see Draine
\& Lazarian 1999) and this poses a question of the degree of its 
polarization. Even moderate polarization can substantially interfere
with the ongoing CMB polarization measurements.

In this review, we summarize the properties of the recently
discovered anomalous emission (see Kogut 1999) 
and discuss why it is inconsistent with free-free
or synchrotron explanations and present two competing
explanations of this emission, namely, the spinning grain model
and the 
magneto-dipole model (section~2). Polarization of microwave foreground due
to alignment of ultrasmall grains is covered in section~3, where
we show that a new
solid state effect termed ``resonance paramagnetic relaxation''
can produce the alignment. Polarization of magneto-dipole emission
and its characteristic signatures are discussed in section~4.

\section{Anomalous Microwave Emission}

Until very recently it has been thought that
there are three major components of the diffuse Galactic foreground:
synchrotron emission, free-free radiation from plasma (thermal bremsstrahlung)
and thermal emission
from dust. In the microwave range of 10-90~GHz the latter 
is definitely subdominant, leaving essentially two
components. However, it is exactly in this range that an anomalous
emission was reported (Kogut et al. 1996a, 1996b). In the recent
paper by de Oliveira-Costa et al. (2000) this emission was nicknamed
``Foreground X'', which properly reflects its mysterious nature.
This component is spatially correlated with 100 $\mu$m thermal
dust emission, but its intensity is much higher than one can expect
by directly extrapolating thermal dust emission spectrum
to the microwave range. 

Since its discovery the Foreground X has been detected in the data
sets from Saskatoon (de Oliveira-Costa et al. 1997), OVRO (Leitch et
al. 1997), the 19~GHz survey (de Oliveira-Costa et al. 1998), and
Tenerife (de Oliveira-Costa et al. 1999, Mukherjee et al. 2000).
Present in the range of 10-90~GHz the Foreground X is particularly
disturbing as many CMB experiments are performed or planned
for this frequency range. Thus it is not surprising that the nature
and properties of the Foreground X have been of considerable interest.

Initially, the anomalous emission was identified as thermal  bremsstrahlung
from ionized gas correlated with dust (Kogut et al. 1996a) and presumably produced
by photoionized cloud rims (McCullough et al. 1999). This idea was
subjected to scrutiny in Draine \& Lazarian (1997) and criticized on
energetic grounds. Additional arguments against the free-free hypothesis
became available through correlating anomalous emission with ROSAT X-ray
C Band (Finkbeiner \& Schlegel 1999) and H$\alpha$ with 100~$\mu$m
emission (McCullough et al. 1999). They are summarized in Draine
\& Lazarian (1999). In a recent preprint 
de Oliveira-Costa et al. (2000) used Wisconsin H-Alpha Mapper (WHAM)
survey data and established that the free-free emission ``is about an order
of magnitude below Foreground X over the entire range of frequencies
and latitudes where is detected''. The authors conclude that the
Foreground X cannot be explained as free-free emission. 

The spectrum of the Foreground X is not consistent with
synchrotron emission, and maps at 408 MHz (Haslam 1981) and 
1.42 GHz (Reich \& Reich 1988)
do not correlate with the observed 15-100 GHz intensity,
so the anomalous emission is evidently not synchrotron radiation from
relativistic electrons.

\subsection{Microwave Emission from Spinning Dust}
  
The surprisingly strong correlation of the Foreground X with
100~$\mu$m emission from dust induced Draine \& Lazarian 
(1998a, henceforth DL98a) to
conjecture that this radiation can be related to a particular component of
dust, namely, ultrasmall grains. The existence of such grains follows
from observations by IRAS, COBE-DIRBE and IRTS at wavelength less
than $50~\mu$m. For instance, emission from diffuse clouds at 12
and 25 $\mu$m (Boulanger \& Perault 1988) is believed to be thermal
emission from grains so small that absorption of a single
photon can raise the grain temperature to $\sim 150$~K for the
$25~\mu$m emission and to $\sim 300$~K for the $12~\mu$m emission. Such
grains contain $10^2-10^3$ atoms and must be sufficiently numerous
to account for $\sim 35\%$ of total absorption of energy from
starlight. These grains are usually assumed to be primarily carbonacious
with $\sim 5\% - 10\%$ of the carbon in the interstellar
medium (in the model by Desert, Boulanger \& Puget 1990 polycyclic
aromatic hydrocarbon molecules, PAH, contain $9\%$ of carbon).
Li and Draine (2001) show that the observd IR emission can be reproduced
by a grain model with $\sim 10$\% of the carbon in grains with less
than 500 c atoms. 

How can these ultrasmall grains be responsible for the microwave
emissivity? DL98a appealed to rotational emission
that must emerge when a grain with a dipole moment $\mu$  rotates
with angular velocity $\omega$. The corresponding dipole emission
is proportional to $\mu^2 \omega^4/c^3$ times the number of emitting
grains $n_g$ along the line of sight. Thus to estimate emissivity
Draine \& Lazarian (1997) had to calculate $n_g$, $\omega$ and $\mu$.
An important finding of DL98a and Draine \& Lazarian (1998b, henceforth
DL98b) was that for
reasonable parameter values the rotational radiation from ultrasmall
grains can account for the observed anomalous emission.

%We note, that the very idea of grain rotational emission was first
%discussed by Erickson (1957). However, in the absence of evidence for
%tiny grains, this idea was not pursued further. More recently,
%after the discovery of the population of ultrasmall grains,
%Ferrara \& Dettmar (1994) noted that the rotational emission from such
%grains may be observable. The gist of DL98a,b
%work was to provide a quantitative description of such emission
%and to compare it with observations. Unlike earlier studies, DL98a,b
% did not assume that grains rotate with Brownian
%velocities corresponding to the gas temperature. On the contrary,
% they presented an exhaustive study of different physical processes
%that can excite and damp grain rotation. Those included collisions
%with ions, plasma interactions, emission of infrared photons and
%microwave drag. In fact, the grain velocities were shown to be
%very non-thermal. For instance, for a $10^{-7}$~cm grain Ferrara
%\& Dettmar (1994) estimate grain velocity in warm ionized media
%is $\sim 50$~GHz, while DL98b get for such
%a grain just $8.5$~GHz. Since the emitted power scales as $\omega^4$
%this is a significant difference.  

The DL98a model of anomalous emission is often referred
to as ``emission of spinning charged grains'', which misrepresents
the process.
In general, centroid of grain charge and centroid of grain inertia
do not coincide and indeed, this results in dipole moment $\mu_{charge}$
appearance. However, more significant for grains is the intrinsic
electric dipole moment $\mu_{int}$. The latter originates naturally as
the tiny grains we deal with are essentially large molecules and many of
molecules are known to exhibit polarization of atomic bonds and therefore
intrinsic dipole moments. Of course, highly symmetric molecules are not
expected to have dipole moment. However, under interstellar conditions
we do not expect the carbon skeleton for very small grains to be fully
hydrogenated (Omont 1986). Detachment of hydrogen atoms under UV flux
is expected to render dipole moment to symmetric species, i.e. C$_{24}$H$_{12}$
(coronene).   

How many small grains do we expect to observe along a line of sight?
While the size distribution of tiny grains is still undetermined, it
is quite clear that the number of such grains exceeds the value that one
can obtain by extrapolating the power law distribution obtained 
for much larger grains (see Weingartner \& Draine 2000). 
DL98a considered a lognormal
distribution of tiny grains to estimate $n_g$ and by making use of the 
observed correlation of 100~$\mu$m emission with 21~cm emission,
$I_{\nu}(100~\mu{\rm m})\approx 0.85$ 
MJy sr$^{-1}$ (N$_H$/$10^{20}$~cm$^{-2}$) (Boulanger \& Perault 1988)
infered the excess microwave emission per H atom for a number of models.
Those differed in cosmic carbon abundance in lognormal component,
grain shape and intrinsic dipole moment. For the model with the most
likely set of parameters, DL98a obtained a reasonable fit with observations
available at that time. It is extremely important that new data points
obtained later (de Oliveira-Costa et al. 1998, 
de Oliveira-Costa et al. 1999) correspond to the already published 
model. The observed flattening of the spectrum and its turnover
around $20$~GHz agreed well with the spinning dust predictions.

How else can DL98a,b theory be verified? DL98a predicted
correlations of Foreground X and diffuse 12 $\mu$m emission, while
in DL98b expected microwave emissivities for
 various regions, including reflection
nebulae, photodissociation regions and dark clouds, were calculated.
Finkbeiner et al. (2000) used the Green Bank 140$\prime$
telescope to measure 10~GHz
emission from IRAS dust filaments. Their results provided only upper
limits, while DL98b model predicts 8~$\sigma$ detection. This
stimulated de Oliveira-Costa et al. (2000) to infer that the correlation
of ultrasmall and large (``classical'') dust grain components may 
not be good at small scales. The authors refer to possible grain
separation processes (see Weingartner \& Draine 2000)
that are likely to act over small scales. We note here that grain coagulation
is another process that is likely to be important on small scales within
dense clouds and this process can substantially deplete the small grain
population. de Oliveira-Costa et al. (2000) tested correlations of
Foreground X with
60~$\mu$m map and obtained a marginally better correlation than with
100~$\mu$m map, which is consistent with DL98a predictions. The authors
correctly point out that the crucial test of the spinning dust model
will come when higher
resolution microwave maps, i.e. those by NASA MAP satellite,
become available. It also seems that the Weingartner \& Draine (2000) 
starlight segregation process can be tested by studying correlation of 
100~$\mu$m emission and Foreground X along and perpendicular to magnetic
field lines, which may be determined via optical and infrared
polarimetry. Indeed, the starlight segregation of small and large
grains should happen along magnetic field lines, while grain electric
charge will substantially impede segregation perpendicular to field
lines. 

The expected relative contribution of various foregrounds on 1 degree
scale for intermediate galactic latitudes is shown in Fig.~1. 
It is easy to see that the foreground
signal is minimal at 60-120~GHz.

\begin{figure}
%\plotone{lazarian_1.eps}
\plotfiddle{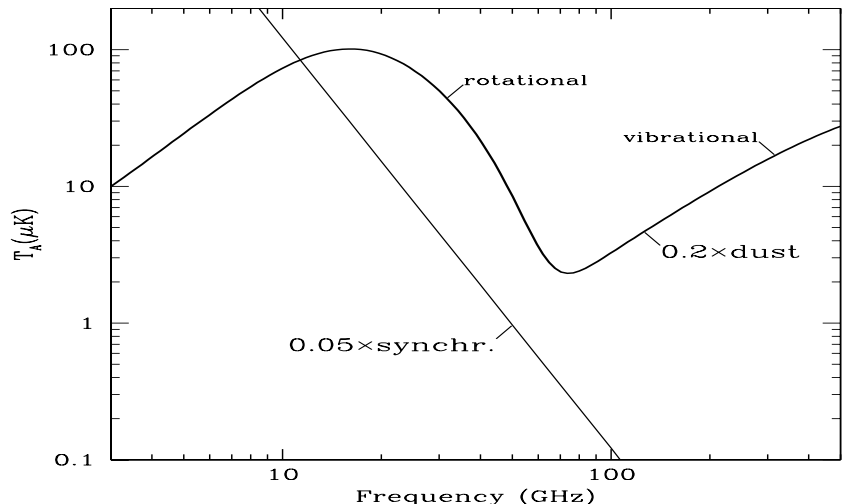}{2.2in}{0}{55}{33}{-160}{-60}
\caption{Estimate of rms variations from Galactic foregrounds
including rotating grains near North Celestial
Pole (from DL98a.). If magneto-dipole emission is responsible for the
Foreground X, the plot may be also used to get a crude idea of how much
of such emission is expected. Details of magneto-dipole emission are
less well constrained due to uncertainties related to magnetic properties
of grain material.}
\end{figure}

\subsection{Microwave Emission from Magnetic Grains}

While the spinning grain hypothesis got recognition in the 
community, the magnetic dipole emission model suggested by Draine
\& Lazarian (1999, henceforth DL99) was left essentially unnoticed. 
This is unfortunate, as magnetic dipole emission provides a
possible alternative explanation to the Foreground X. Magnetic 
dipole emission is negligible at optical and infrared frequencies. 
However, when the
frequency of the oscillating magnetic field approaches the precession
frequency of electron spin in the field of its neighbors, i.e.
$10$~GHz, the magneto dipole emissivity becomes substantial.

How likely is that grains are strongly magnetic? Iron is the fifth
most abundant element by mass and it is well known that it resides in
dust grains (see Savage \& Sembach 1996). If $30\%$ of grain mass is
carbonaceous, Fe and Ni contribute approximately $30\%$ of the remaining
grain mass. Magnetic inclusions are widely discussed in grain
alignment literature (Jones \& Spitzer 1967, Mathis 1986, Martin 1995, 
Goodman \&
Whittet 1996). 
If a substantial part of this material is ferromagnetic
or ferrimagnetic, the magneto-dipole emission can be comparable to that
of spinning grains. Indeed, calculations in DL99 showed that less than
$5\%$ of interstellar Fe in the form of metallic grains or inclusions
is necessary to account for the Foreground X at 90~GHz, while magnetite,
i.e. Fe$_3$O$_4$, 
can account for a considerable part of the anomalous emissivity over
the whole range of frequencies from 10 to 90~GHz. Adjusting the
magnetic response of the material, i.e. making it more strongly magnetic
than magnetite, but less magnetic than pure metallic Fe, it is possible
to get a good fit for the Foreground X (DL99).

How can magneto-dipole emission be distinguished from that from
 spinning grains?
The most straightforward way is to study microwave emission from regions
of different density. The population of small grains is 
depleted in dark clouds (Leger and Puget 1984) 
and this should result in a decrease of
contribution from spinning grains.
%Private communication from Dick Crutcher who attempted such measurements
%corresponds to this tendency, but the very detection of microwave
%emissivity is a 3$\sigma$ result. 
Obviously the corresponding  measurements are 
highly desirable. 
%Marginally better correlation of Foreground X 
%with 60 $\mu$m compared to its correlation with 100~$\mu$m also favors
%spinning dust, but the result is not definitive either. 
As for now,
magnetic grains remain a strong candidate process for producing
part or even all of Foreground X. 
%In any case, even if magnetic
%grains provide subdominant contribution, this can be important
%for particular cases of CMB and interstellar studies. For instance,
%polarization from magnetic grains may dominate that from spinning
%grains even if the emission from spinning grains is more of higher
%level. 

\section{Polarization of Spinning Grain Emission}

Microwave emission from spinning grains is expected to be polarized if
grains are aligned. Alignment of ultrasmall grains which are
essentially large molecules is likely to be different from alignment 
of large (i.e. $a>10^{-6}$~cm) grains for which the
theory of grain alignment (see review by Lazarian 2000) has been developed.

One of the mechanisms that might produce alignment of the ultrasmall
grains is the
paramagnetic dissipation mechanism suggested half a century ago
by Davis and Greenstein (1951) as a means of explaining the
polarization of starlight. The Davis-Greenstein alignment mechanism is
straightforward: for a spinning grain
the component of interstellar magnetic field
perpendicular to the grain angular velocity varies in grain coordinates,
resulting in time-dependent magnetization, associated
energy dissipation, and a torque acting on the grain.
As a result grains  tend to rotate
with angular momenta parallel to the 
interstellar magnetic field. 
%Although recent research (Draine \& Weingartner 1996, 1997, Lazarian \& Draine
%1999a,b) suggests that paramagnetic alignment may not be the dominant 
%alignment 
%mechanism for it 
%may be promising for small ($a\leq 10^{-6}$~cm) grains. 

Are the ultrasmall grains paramagnetic? The answer to this question
is positive owing to the presence of free radicals, paramagnetic
carbon rings  (see Altshuler
\& Kozyrev 1964) and captured ions.
For paramagnetic grains, the alignment time-scale
 $\tau\approx
10^4$~yr~$ (a/10^{-6}~{\rm cm} )^2 (10^{-13}~{\rm s}/K)$ with
$K(\omega)\equiv {\rm Im}(\chi)/\omega$, where $\omega$ is the 
angular rotational
velocity and $\chi(\omega)$ is the magnetic susceptibility.
The characteristic time of grain magnetic response is
the electron precession time in the field of its neighbors, which
is also called spin-spin relaxation time and is denoted $\tau_2$.
If $\omega \leq \tau_2^{-1}\sim
10^8$~s$^{-1}$, normal materials at $T\approx 20$~K
have $K\approx 10^{-13}$~s. For higher frequencies , however, $K(\omega)$
begins to decrease rapidly (DL99). 
As discussed earlier spinning grains must rotate
much faster to account for the Foreground X, thus apparently calling into 
question the efficacy of alignment by paramagnetic dissipation.

Lazarian \& Draine (2000, henceforth LD00) found 
that the traditional picture
of paramagnetic relaxation is 
incomplete, since it
disregards the splitting of energy levels within a rotating
body. 
Unpaired
electrons spin parallel and antiparallel
to the grain angular velocity have different energies causing
 the so-called ``Barnett magnetization'' (Landau \& Lifshitz 1960).
The Barnett effect, the inverse of the Einstein-De Haas effect, 
consists of the spontaneous magnetization of
a paramagnetic body
rotating in field-free space. This effect can be understood in
terms of the lattice sharing part of its angular momentum with 
the spin system. Therefore the implicit assumption in Davis 
\& Greenstein (1951)-- 
that the magnetization within a {\it rotating grain} in a {\it static} 
magnetic field is equivalent to the magnetization within a 
{\it stationary grain} in a {\it rotating} magnetic field --
is clearly not exact.

If electrons within a rotating grain
are treated as nearly free, the magnetization of
the grain is the same for a stationary grain in a ``Barnett-
equivalent'' magnetic field directed along the grain angular 
velocity $\omega$
and having an amplitude 
$
H_{\rm BE}=\hbar\omega/(g\mu_{\rm B})
$, where $g$
is the electron gyromagnetic ratio $\approx 2$ and $\mu_{\rm B}$ is the Bohr magneton.
In these conditions the component of magnetic field perpendicular 
to $ \omega$ causes electron
spin resonance (see Athertson 1973).

LD00 called the process of paramagnetic relaxation within
a rotating body ``resonance relaxation'' as opposed to Davis-Greenstein
relaxation that disregards the spontaneous magnetization of a rotating body.
Solving the 
Bloch equations (Bloch 1946) LD00 obtained the  following expression
for
the imaginary part of the grain paramagnetic susceptibility (the
part responsible for dissipation and therefore alignment):
\begin{equation}
%\chi^{\prime\prime}=\chi_0 
Im(\chi)=\chi_0 
\frac {\omega \tau_2}
{1+\gamma^2 g^2 \tau_1 \tau_2H_1^2}~~~,
\label{eq:chi''}
\end{equation}
where  $\gamma\equiv e/2 m_e c=8.8\times 10^6$~s$^{-1}$G$^{-1}$,
$\tau_1$ is the spin-lattice relaxation time, and $H_1$ is
interstellar magnetic field intensity.
Unlike the corresponding expression in Davis-Greenstein theory,  
eq.~(1) does not vanish for $\omega$ much larger
than the spin-spin relaxation time $\tau_2$. The saturation,
however, depends on the value of $H_1$
and $\tau_1$. The latter parameter
was calculated in LD00 using Raman scattering of phonons, but
the calculations are based on the so-called Waller theory, which
is known to overestimate $\tau_1$ considerably. Thus laboratory 
measurements of relaxation within isolated grains are required.

\begin{figure}
%\plotone{lazarian_2.ps}
\plotfiddle{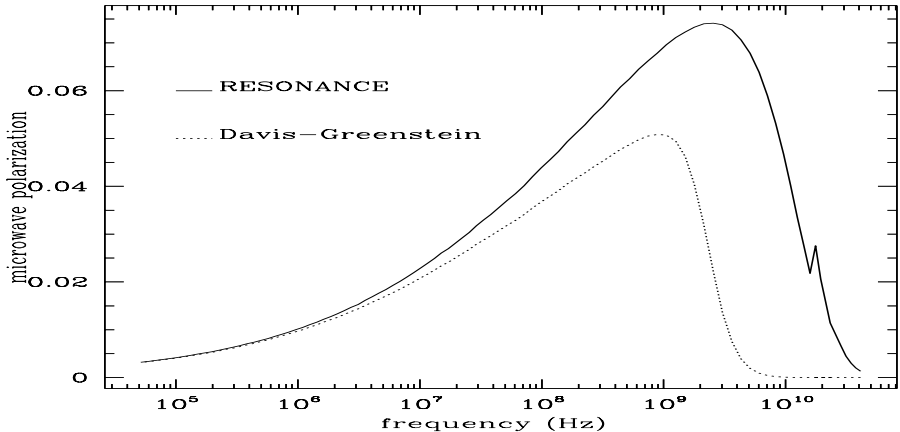}{2.0in}{0}{62}{30}{-180}{-55}
\caption{Measure of grain alignment for both
	resonance relaxation and Davis-Greenstein relaxation for grains in
	the cold interstellar medium as a function of frequency (from LD00).
For resonance relaxation the saturation effects 
(see eq.\ (1)) are neglected, which means
that the upper curves correspond to the {\it maximal} values allowed by the 
paramagnetic mechanism.}
\end{figure}

Fig.~2 shows the predictions of the resonance relaxation mechanism
for cold interstellar gas
assuming that the spin-lattice relaxation is fast.  The discontinuity 
at $\sim 20$~GHz
is due to the assumption that smaller grains are planar, and larger
grains are spherical. The microwave emission will be polarized
in the plane perpendicular to magnetic field. 
The dipole rotational 
emission predicted in DL98a,b is sufficiently 
strong that polarization of a few percent
may interfere with efforts to measure the polarization of
the CMB.

Can we check the alignment of ultrasmall grains via infrared polarimetry? The answer to this
question is ``probably not''. Indeed, as discussed earlier,
infrared emission from ultrasmall grains, e.g. 12 $\mu$m emission,
takes place as grains absorb UV photons. These photons raise
grain temperature, randomizing grain axes in relation to
its angular momentum (see Lazarian \& Roberge 1997). Taking values
for Barnett relaxation from Lazarian \& Draine (1999), we get
the randomization time of the $10^{-7}$~cm grain to be
 $2\times 10^{-6}$~s, which is less than grain cooling time. As the
result, the emanating infrared radiation will be polarized very marginally.
If, however, Barnett relaxation is suppressed, the randomization time 
is determined
by inelastic relaxation (Lazarian \& Efroimsky 1999) and is 
$\sim 0.1$~s, which would entail a partial polarization of
infrared emission. 

\section{Polarization of Magneto-Dipole Emission}

The mechanims of producing polarized
magneto-dipole emission is similar to that
producing polarization of electro-dipole  thermal emission 
emitted from aligned non-spherical grains (see Hildebrand 1988).
There are two
significant differences, however. First, strongly magnetic
grains can contain just a single magnetic domain. Further magnetization
along the axis of this domain is not possible and therefore the 
magnetic permeability of the grains gets anisotropic: $\mu=1$ along
the domain axis, and $\mu=\mu_{\bot}$ for a perpendicular direction.
Second, even if a grain contains tiny magnetic inclusions and can be
characterized by isotropic permeability, polarization that it produces
is orthogonal to the electrodipole radiation emanating through
electro-dipole vibrational emission. In case
of the electo-dipole emission, the longer grain axis defines the vector
of the electric field, while it defines the vector of the 
magnetic field in case
of magneto-dipole emission.

The results of calculations for single domain iron particle (longer axis
coincides with the domain axis) and a grain with metallic Fe inclusions
are shown in Fig.~3. Grains are approximated by ellipsoids $a_1<a_2<a_3$
with ${\bf a_1}$ perfectly aligned 
parallel to the interstellar magnetic field ${\bf B}$. The polarization
is taken to be positive when the electric vector of emitted radiation
is perpendicular to ${\bf B}$; the latter is the case for electro-dipole
radiation of aligned grains. This is also true (see Fig.~3) for high
frequency radiation from single dipole grains. It is easy to see
why this happens. For high frequencies $|\mu_{\bot}-1|^2\ll 1$
and grain shape factors are unimportant. The only important thing is
that the magnetic fluctuations happen perpendicular to ${\bf a_1}$. 
With ${\bf a_1}$ parallel to ${\bf B}$, the electric fluctuations
tend to be perpendicular to ${\bf B}$ which explains the polarization
of single domain grain being positive. For lower frequencies magnetic
fluctuations tend to happen parallel to the intermediate size axis 
${\bf a_2}$. As the grain rotates about ${\bf a_1}\|{\bf B}$,
the intensity in a given direction reaches maximum when an observer
sees the ${\bf a_1} {\bf a_2}$ grain cross section. Applying earlier
arguments it is easy to see that magnetic fluctuations are parallel
to ${\bf a_2}$ and therefore for sufficiently large $a_2/a_1$ ratio
the polarization is negative. {\it 
The variation of the polarization direction with
frequency presents the characteristic signature of magneto-dipole emission
from aligned single-dipole grains and it can be used to separate this
component from the CMB signal}. Note that the degree of polarization is
large, and such grains may substantially interfere with the attempts
of CMB polarimetry. Even if the
intensity of magneto-dipole emission is subdominant
to that from rotating grains, it can still be quite important in
terms of polarization.
A relatively weak polarization response is expected for grains with
magnetic inclusions (see Fig.~3). The resulting emission is negative
as magnetic fluctuations are stronger along longer grain axes, while
the short axis is aligned with ${\bf B}$.

\begin{figure}
\plotfiddle{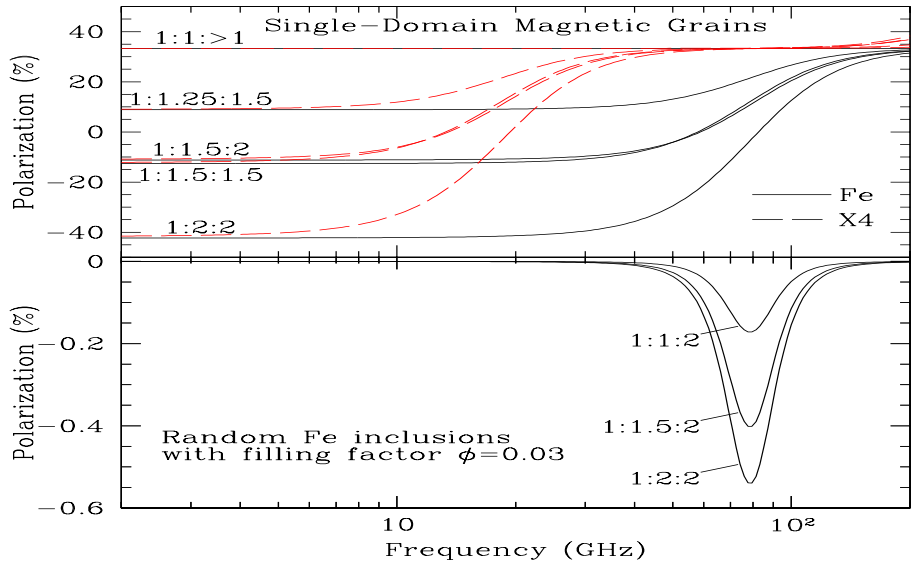}{2.5in}{0}{60}{37}{-180}{-60}
\caption{ Polarization from magnetic grains (from DL99). Upper panel:
Polarization of thermal emission from perfectly aligned single
domain grains of metallic Fe (solid lines) or hypothetical magnetic
material that can account for the Foreground X (broken lines).
Lower panel: Polarization from perfectly aligned grains with
Fe inclusions (filling factor is 0.03). Grains are ellipsoidal and
the result are shown for various axial ratios.
}
\end{figure} 

\section{Summary}

The principal points discussed above are as follows:

Dust emission in the microwave range is stronger than it was
thought to be. 
Ultrasmall grains, whose existence is required to explain
$\lambda<50\mu$m emission, should also produce microwave emission as
they rotate rapidly. Moreover, magnetic fluctuations within large
interstellar grains constitute another important source of microwave
dust emissivity.

Both dipole rotational emission from ultrasmall grains and magneto-dipole
emission may explain the 10-100~GHz anomalous emission correlated with
interstellar dust. At the moment, rotating grains seem to be more favored
candidates to account for the anomalous emission, but further tests
that include measurements from e.g. dark globules are necessary.

Microwave emission of both origins is expected to be partially polarized.
``Resonance paramagetic relaxation'' is the process that can enable
alignment and therefore polarization from ultrasmall grains. Although
the details of the process still require laboratory testing, it looks
that the polarization is marginal beyond 40~GHz. On the contrary,
magneto-dipole emission may be
substantially polarized for higher frequencies.
Thus it can be important in terms of polarization even if it is
subdominant in terms of total emissivity.   
   
{\bf Acknowledgments.} It is a pleasant duty to thank D. Spergel for
attracting our attention to the problem. This review summarizes our
``microwave'' research done with Bruce Draine and I am extremely grateful
to him for great time of working together. Comments by Bruce Draine
and John Mathis improved the manuscript.

\end{document}